\documentclass{article}
\usepackage{arxiv}
\usepackage{url} 
\usepackage{lineno}
\usepackage{soul}
\usepackage{amsmath} 
\usepackage{amssymb} 
\usepackage{graphicx} 
\usepackage{caption} 
\usepackage{hyperref}       
\usepackage{url}            
\usepackage{natbib}
\usepackage{doi}

\title{Predictability Limit of the 2021 Pacific Northwest Heatwave from Deep-Learning Sensitivity Analysis}

\newif\ifuniqueAffiliation

\ifuniqueAffiliation 
\author{ \href{https://orcid.org/0000-0000-0000-0000}{\includegraphics[scale=0.06]{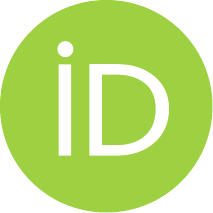}\hspace{1mm}David S.~Hippocampus}\thanks{Use footnote for providing further
		information about author (webpage, alternative
		address)---\emph{not} for acknowledging funding agencies.} \\
	Department of Computer Science\\
	Cranberry-Lemon University\\
	Pittsburgh, PA 15213 \\
	\texttt{hippo@cs.cranberry-lemon.edu} \\
	\And
	\href{https://orcid.org/0000-0000-0000-0000}{\includegraphics[scale=0.06]{orcid.pdf}\hspace{1mm}Elias D.~Striatum} \\
	Department of Electrical Engineering\\
	Mount-Sheikh University\\
	Santa Narimana, Levand \\
	\texttt{stariate@ee.mount-sheikh.edu} \\
}
\else
\usepackage{authblk}

\newbox{\orcid}\sbox{\orcid}{\includegraphics[scale=0.06]{orcid.pdf}} 
\author[1,2]{%
	\href{https://orcid.org/0009-0006-7242-2351}{\usebox{\orcid}\hspace{1mm}P. Trent Vonich \thanks{\texttt{tvonich@uw.edu}}}%
}
\author[1]{%
	\href{https://orcid.org/0000-0001-8486-9739}{\usebox{\orcid}\hspace{1mm}Gregory J. Hakim \thanks{\texttt{ghakim@uw.edu}}}%
}
\affil[1]{Department of Atmospheric Sciences, University of Washington, Seattle, WA, USA}
\affil[2]{Air Force Institute of Technology, Wright-Patterson AFB, OH, USA}
\fi

\renewcommand{\shorttitle}{Atmospheric Predictability from Deep-Learning Sensitivity Analysis}

\begin{document}
\maketitle



%
%

\begin{abstract}
The traditional method for estimating weather forecast sensitivity to initial conditions uses adjoint models, which are limited to short lead times due to linearization around a control forecast. The advent of deep-learning frameworks enables a new approach using backpropagation and gradient descent to iteratively optimize initial conditions to minimize forecast errors. We apply this approach to forecasts of the June 2021 Pacific Northwest heatwave using the GraphCast model, yielding over 90\% reduction in 10-day forecast errors over the Pacific Northwest. Similar improvements are found for Pangu-Weather model forecasts initialized with the GraphCast-derived optimal, suggesting that model error is not an important part of the initial perturbations. Eliminating small scales from the initial perturbations also yields similar forecast improvements. Extending the length of the optimization window, we find forecast improvement to about 23 days, suggesting atmospheric predictability at the upper end of recent estimates.
\end{abstract}

\section{Introduction}

Atmospheric predictability is limited by sensitivity to the initial conditions \citep[e.g.,][]{lorenz1996predictability,bauer2015quiet}, and analysis of operational forecasts suggest that this limit is currently around 10 days \citep{zhang2019predictability}, but may be extended using ensemble prediction to 16--23 days \citep[e.g.,][]{buizza2015forecast}. For any individual case, forecast sensitivity can be quantified by defining a forecast metric to measure errors, and then computing the relationship between the metric and the initial conditions. Using a metric based on forecast error, for example, one can compute the changes to the initial condition that reduce forecast error. The most common approach to this problem employs adjoint models \citep[e.g.,][]{errico1997adjoint,langland1995evaluation,doyle2019adjoint}, which linearize perturbations about a control forecast, limiting the lead time over which the sensitivity can be determined to less than about 5 days. A similar approach is available using automatic differentiation in deep-learning frameworks to identify the sensitivity of outputs to inputs \citep{gagne2019interpretable,mcgovern2019making,toms2020physically}. There are several advantages to the deep-learning approach, including a massive increase in computational efficiency relative to physical models, which enables deep searches for optimal initial conditions using nonlinear gradient descent. We apply this method to the extreme heatwave of June 2021 over the Pacific Northwest (PNW) region of North America using the GraphCast \citep{lam2023learning} deep learning model and the JAX framework for computing sensitivity using automatic differentiation, with backpropagation and gradient descent optimization to identify the initial conditions that produce the best 10 day forecast of the heatwave.

There are several aspects of the deep-learning method that distinguish it from adjoint methods for finding initial conditions that improve forecasts. In the adjoint method, the error in the forecast is propagated backward in time to the initial condition using the adjoint version of the forecast model. The adjoint model is the transpose of the linearized version of the forecast model about the control forecast, where the transpose operation reverses the role of inputs and outputs (and the order of time; hence backward propagation). The adjoint model involves making choices for how to handle on/off processes, and involves explicit coding, which is tedious and expensive. In contrast, in deep learning frameworks the models consist of layers with linear operations and analytic nonlinear activation functions between layers. As a consequence, derivatives simply apply the chain rule on known linear and nonlinear functions, which is well suited for automatic differentiation. Backpropagation of gradients computed in this way is used to train deep-learning models by adjusting model weight parameters to better fit the output to training data. Here we hold the model weights constant, and backpropagate forecast error to adjust the initial condition in order to better fit the verification data in the forecast, similar to the approach of \citet{toms2020physically}. Another important difference with the deep-learning approach is optimization, which uses a combination of the forecast model for the forward (nonlinear) pass and backpropagation of gradients with respect to the error on the output layer. The fact that the forecast model is extremely computationally inexpensive is essential to this approach. With adjoint approaches, typically only a single pass is made rather than an iteration due to the large expense of running the fully nonlinear physics model on the forward pass. Nevertheless, optimization strategies similar in spirit to the backpropagation approach have been proposed using the adjoint technique with physics models \citep[e.g.,][]{mu2000nonlinear,mu2003conditional}.

During the past two years a number of deep-learning models have emerged with forecast skill comparable to operational models \citep{rasp2023weatherbench}. These models are typically trained on the ERA5 reanalysis dataset \citep{hersbach2020era5} to produce the best forecasts on timescales up to about 10 days. In addition to highly skillful forecasts on data outside the training set, the models also appear to have encoded aspects of the basic laws of physics into their parameters as suggested by the idealized dynamical experiments of \citet{hakim2024dynamical}. However, this generation of models is also lacking in power in forecasting smaller scales \citep{bonavita2023limitations} and exhibit different forecast error growth from physics models at small amplitude \citep{selz2023can}. Because deep-learning models are trained to capture predictable signals, they inevitably filter out unpredictable signals, which leads to smoothing somewhat analogous to ensemble averaging \citep{bonavita2023limitations}. Our sensitivity method will inherit these deficiencies, although the algorithm is general and future applications with improved models will likely overcome the weaknesses of the current generation of deep-learning models.

We select the PNW June 2021 heatwave for an example illustration of the method for two reasons. First, it is an exceptional heatwave, exceeded globally by only five other events since 1960 \citep{thompson20222021}, and is not included in the training data for the models we use. The event was due to an exceptional blocking anticyclone in the upper troposphere that was affected by latent heating from water vapor transported across the North Pacific ocean the preceding week \citep{lin20222021,schumacher2022drivers}. Seasonal timing near the solstice and anomalously dry soil conditions \citep{conrick2023influence} also appear to also play a contributing role. As such the case provides a good test case of deep-learning models in simulating an unseen extreme event with a mixture of contributing physical processes. Second, operational forecasts of the event from the European Centre for Medium Range Weather Forecasts (ECMWF) Integrated Forecast System (IFS) were very skillful in predicting the event at least 7 days in advance \citep{lin20222021,emerton2022predicting}, although the amplitude of the event was underpredicted. At longer lead times, out to about 11 days, the IFS ensemble mean captured the event, but with much weaker amplitude; however, one ensemble member produced an excellent forecast \citep{leach2024heatwave}. The fact that one ensemble member produced a very good forecast suggests that the $\sim$10 day forecasts of this event are potentially improvable by modifying the initial conditions, and we seek to determine to what extent our method can find an optimal initial condition that achieves this objective.

The remained of the paper is organized as follows. In section \ref{sec:model_data} we discuss the  GraphCast model \citep{lam2023learning}, which is used for forecast optimization, and the data used to initialize and verify forecasts. The method for forecast optimization is described in section \ref{sec:methods}, along with the loss function we optimize using backpropagation and gradient descent. Results are presented in section \ref{sec:results}, for both GraphCast and the Pangu-Weather model \citep{bi2023} to test for the importance of model error in the optimal initial conditions. A concluding discussion is provided in section \ref{sec:conlusions}.

\section{Model \&\ Data\label{sec:model_data}}
\subsection{Graphcast}

The GraphCast model \citep{lam2023learning} is available in three versions, differentiated primarily by the resolution of the atmospheric input and output states: 37-levels on a 0.25$^{\circ}$ grid, 13-levels on a 0.25$^{\circ}$ grid, and 13-levels on a 1.0$^{\circ}$ grid. Given the GPU memory requirements of this project, we use the 1.0$^{\circ}$ model (henceforth simply referred to as GraphCast). The model contains 36.7 million parameters and is trained on 1.0$^{\circ}$ ERA5 reanalysis data from 1979 to 2015. It predicts six atmospheric variables (geopotential height, temperature, water vapor specific humidity, vertical velocity, and horizontal wind components) on thirteen pressure levels, four surface variables (mean-sea-level pressure, 2m air temperature, and 10m wind components), and 6-hour accumulated precipitation. Inference involves an input context consisting of two atmospheric states separated by six hours, and outputs the state six hours in the future. The output is fed autoregressively along with the preceding six-hour state as input to continue the forecast indefinitely. For further detail on GraphCast initialization and variables, see Table 1 of \citet{lam2023learning}.

\subsection{Input Data}

Our objective is to reduce forecast error as defined by a loss function, which we take to be the Graphcast loss function used in training (defined in section \ref{sec:methods}). We experimented with a range of lead times for optimization, and settled on the ten-day forecast for the main results since this is around the time that operational forecasts first began to resolve the event. We will also explore a wider range of optimization times for a different initial time to illustrate that the findings are not peculiar to the case considered for the main results. For all experiments we use ERA5 reanalysis data, acquired from the Copernicus Data Storage (CDS) for the time period 00 UTC 1 June 2021 to 00Z 30 June 2021, and downsampled to the 1$^{\circ}$ grid required by the version of GraphCast that we use. Accumulated precipitation is only available at one-hour intervals from CDS. We replicate the method of \citet{lam2023learning} and sum over the preceding six hours of accumulated precipitation; e.g., accumulated precipitation at 00 UTC is the sum of the one-hour values from 19 to 00 UTC. This summation is extended across the entire month of June to complete our dataset. 

GraphCast also requires two static variables, a land-sea mask and geopotential at the surface (elevation). These are included with the open-source distribution of GraphCast on GitHub (\url{https://github.com/google-deepmind/graphcast}). Time-dependent specified variables include top-of-atmosphere solar radiation, time of day, and day of the year, which are generated by the model code. When producing an optimal set of initial conditions, we only perturb the state variables, leaving the static and time-dependent specified input variables unmodified (although we note that it is possible to include these in the optimization).

\section{Methods\label{sec:methods}}
\subsection{Gradient-Based Input Optimization}
Backpropagation, short for backward propagation of errors, is the foundation on which machine learning models are trained and relies on the computation of gradients. Given input state ${\bf x}_i$ for iteration (i.e, epoch) $i$ ($i=1$ represents ERA5 data), we compute an increment based on the gradient of the forecast loss, $\mathcal{L}({\bf N}({\bf x}_i))$ with respect to the inputs, where ${\bf N}$ represents GraphCast inference to the selected forecast optimization time:
\begin{equation}
{\bf x}_{i+1} = {\bf x}_i - \eta\frac{\partial \mathcal{L}}{\partial {\bf x}}_i.
\label{eqn:increment}
\end{equation}
\noindent The derivative in (\ref{eqn:increment}) involves mapping the gradient of the output back through the GraphCast neural network for every time step.
The gradient quantifies the sensitivity of the forecast loss to the input values as for adjoint sensitivity, and the step size $\eta$ is known as the learning rate in a machine learning context. \citet{toms2020physically} illustrate this approach for an El Niño Southern Oscillation (ENSO) classifier model, revealing the sea surface temperature pattern most strongly linked by the model to El Niño conditions. This technique, which they call ``backward optimization,'' originated in image classification tasks \citep{simonyan2013deep,yosinski2015understanding,olah2017feature}. Here we seek the inputs that produce the smallest forecast error over a chosen lead time as defined by the loss. 

\subsection{Optimization Process}

The method used to produce a set of optimized inputs is as follows: 
\begin{enumerate}
    \item Given a set of inputs, produce a forecast at a chosen lead time. Here we initialize the first iteration with an ERA5 reanalysis state. 
    \item Calculate the forecast error relative to a verification state using a prescribed loss function. Here we verify against ERA5 and adopt the loss function used to train Graphcast.
    \item With the model parameters frozen, calculate the gradient of the loss function with respect to the inputs using the open-source JAX framework (available on Github).
    \item Perform gradient-descent by updating the inputs using the loss gradient by equation (\ref{eqn:increment}). We utilize the Adam optimizer from the Python Optax library with a learning rate of \( 10^{-3}\) and default $\beta_1$ value of 0.9. For optimization longer than 14 days we reduce the learning rate to \(10^{-4}\) to allow for smaller steps needed to  navigate the gradient descent process at longer lead times.
    \item Repeat 1)--4) until the loss plateaus or a prescribed maximum iteration is reached. We use a maximum of 100 iterations, and an example of the loss as a function of epoch is shown in Fig.~S1.
\end{enumerate}

\subsection{The Loss Function}
We use the scalar loss used to train Graphcast, which in essence is a weighted mean-squared-difference between the target output and the predicted output averaged across time, variable, and space. For predicted state $\hat{x}$ and verification state $x$, the loss is defined as  
\begin{equation}
\mathcal{L}_{\text{MSE}} = 
\underbrace{
\frac{1}{T_{\text{time}}} \sum_{\tau \in 1:T_{\text{time}}}}_{\text{lead time}}
\underbrace{
\frac{1}{|G_{1.0^\circ}|} \sum_{i \in G_{1.0^\circ}}
}_{\text{spatial location}}
\underbrace{
\sum_{j \in J}
}_{\text{variable-level}}
s_j w_j a_i \left(\hat{x}_{i,j}^{t_0+\tau} - x_{i,j}^{t_0+\tau}\right)^2.
\end{equation}
\noindent Here, $w$ is a weight by pressure level (maximum at the surface and decreasing upward), $a$ is the grid-cell area, and $s$ is a standardization parameter computed from time-differences in the GraphCast training data. We refer the reader to Section 4.2 of the Materials and Methods section of \citet{lam2023learning} for more details. We optimize for two spatial domains (i.e. $G$), one global, and the other specific to the PNW (42.0$^{\circ}$N to 60.0$^{\circ}$N, 130.0$^{\circ}$W to 110.0$^{\circ}$W). Finally, we note that, due to forecast error growth, the time average is heavily weight toward times closer to the final (validation) time. Moreover, in figures shown subsequently, we plot the loss as a function of time without the time average in order to reveal error changes in time.

\section{Results\label{sec:results}}
\subsection{Ten-Day Regionally-Optimized Heatwave Forecast}

\begin{figure}[ht!]
\begin{center}
  \noindent\includegraphics[width=.85\textwidth]{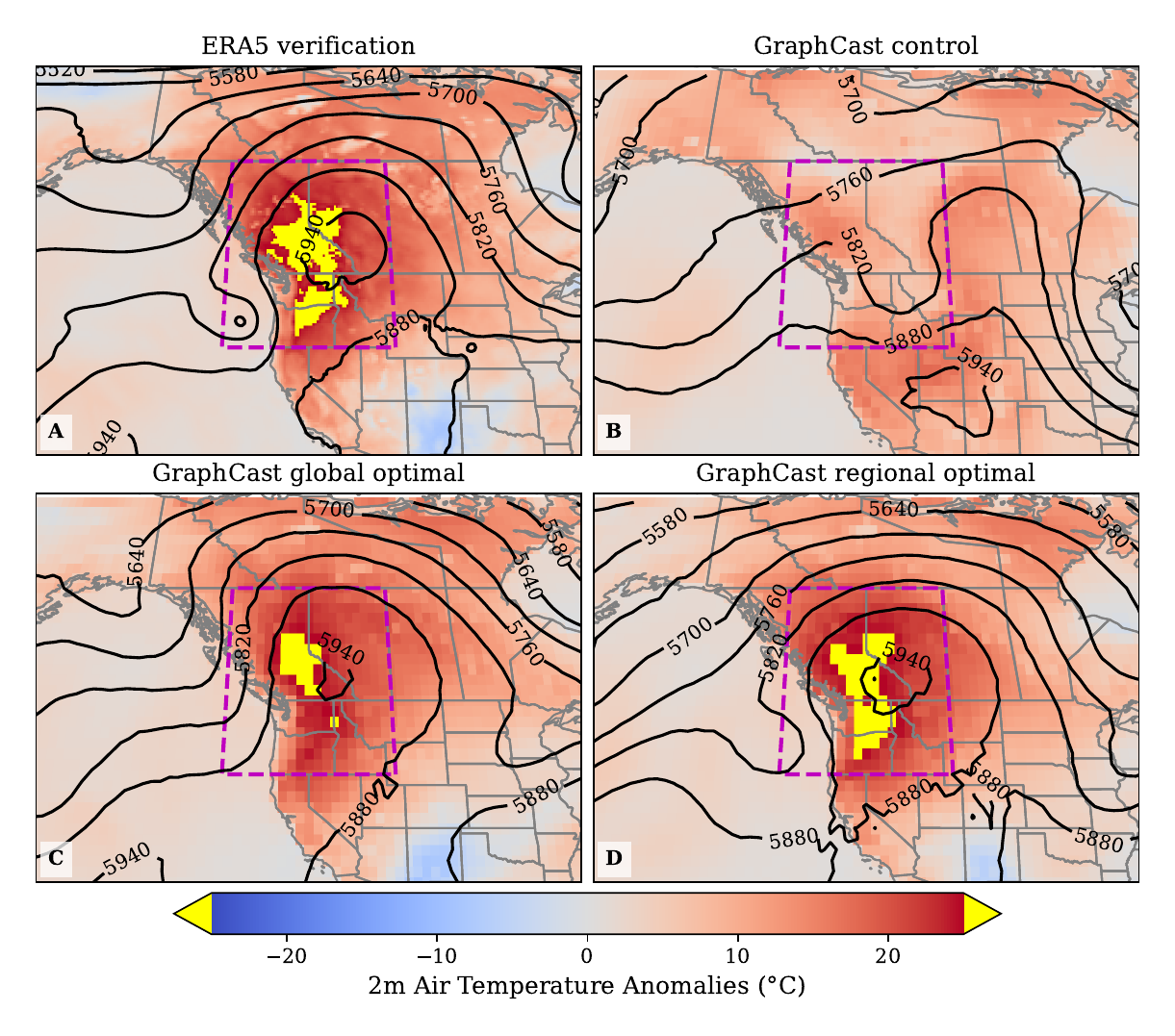}
\end{center}
  \caption{500 hPa geopotential height (contours) and 2m air temperature anomalies (colors) at 00UTC 30 June 2021 for A) ERA5 reanalysis, B) GraphCast 10-day control forecast, C) GraphCast 10-day forecast from globally optimal initial condition, and D) GraphCast 10-day forecast from regionally optimal initial condition. All forecasts are initialized 00UTC 20 June 2021. Anomalies are relative to ERA5 June climatology 1979--2020, with anomalies greater than 25$^{\circ}$C shaded yellow.} \label{fig:graphcast_10day}
\end{figure}

The target ERA5 verification field at 00 UTC 30 June shows conditions at the peak of the PNW heatwave (Fig.~\ref{fig:graphcast_10day}A). At 500 hPa, a large ridge of high geopotential height with maximum values over 5940m is located over the region of greatest 2m temperature anomalies with values in excess of 25$^{\circ}$C from the ERA5 June monthly mean 1979--2020 (yellow shading). A region of low geopotential height is located west of the ridge, with a broad trough in the Gulf of Alaska, and a smaller trough west of the Oregon coast. The control 10-day forecast for GraphCast, started from the ERA5 analysis at 00UTC 20 June, shows a trough over the PNW near the ridge in the verification field (Fig.~\ref{fig:graphcast_10day}B). Control forecast 2m temperature anomalies in the region are slightly above normal, partly reflecting seasonality (30 June verification compared to June climatology), but also the fact that the trough is located in the middle of a larger-scale ridge over western North America. We note the difference in resolution between the GraphCast forecasts (1$^{\circ}$ as compared to ERA5 0.25$^{\circ}$).

There are two results for GraphCast-optimized initial conditions, one that minimizes the loss function globally, and the other over the PNW region (denoted by the magenta dashed lines in Fig.~\ref{fig:graphcast_10day}). Even though the global-optimized initial condition does not specifically target the PNW, the forecast significantly improves upon the control in this region (Fig.~\ref{fig:graphcast_10day}C). The trough over the PNW in the control is replaced by a ridge, similar to the verification field, and 2m air temperature anomalies in excess of 25$^{\circ}$C are located in the same location as in the verification field, at the lower resolution of GraphCast. The regionally optimized solution produces a nearly perfect 10-day forecast over the PNW, down to the details of the ridge and trough at 500 hPa, and the amplitude and structure of the 2m temperature anomalies (Fig.~\ref{fig:graphcast_10day}D).

\begin{figure}[ht!]
\begin{center}
  \noindent\includegraphics[width=.85\textwidth]{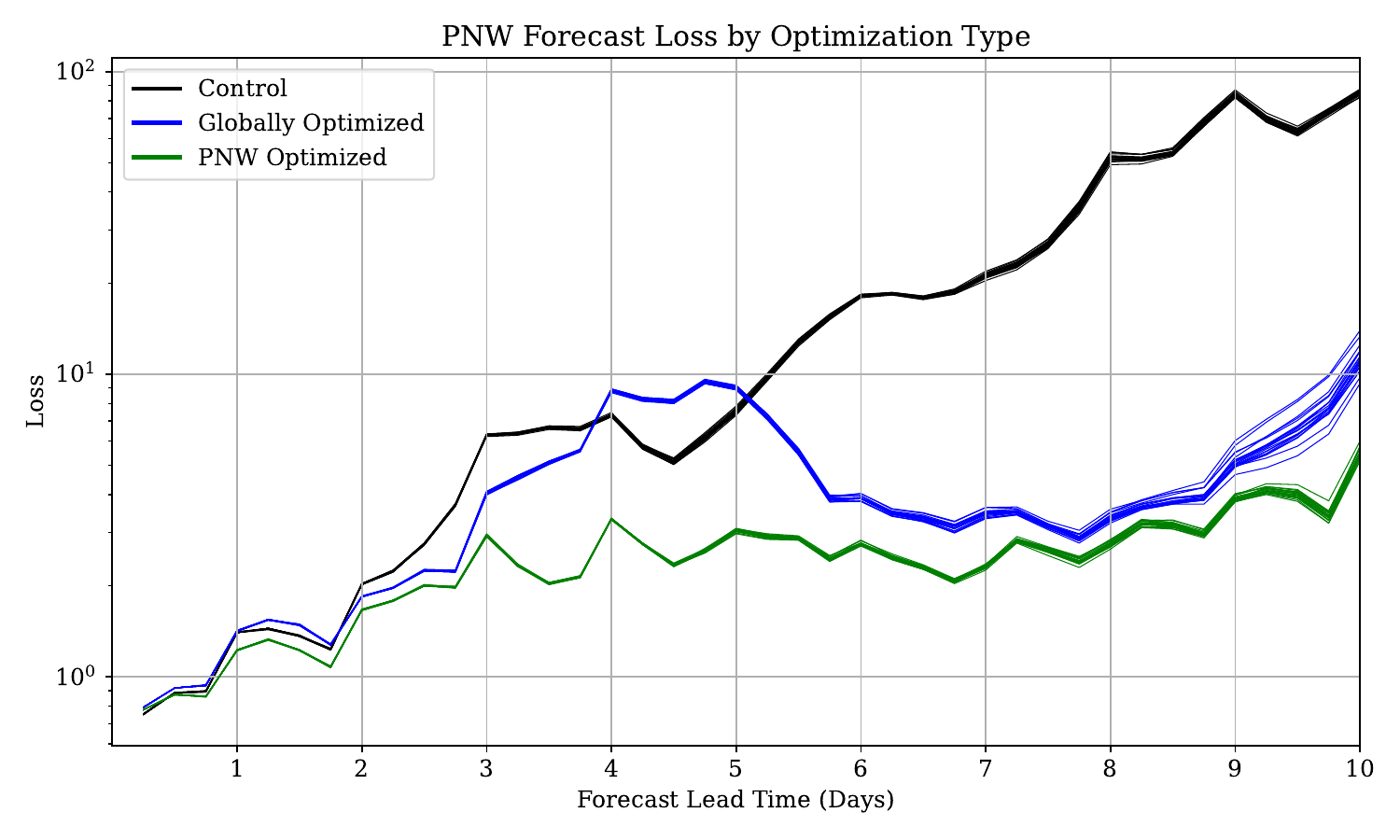}
\end{center}
  \caption{GraphCast instantaneous loss (i.e, not time averaged) on the PNW domain as a function of time. The GraphCast control forecast, initialized with the ERA5, is shown in black, the forecast from the initial condition that minimizes the global loss in blue, and the forecast that minimizes the loss on the PNW domain in green.  A 20-member ensemble is shown for each experiment representing solutions from identical initial conditions that  diverge due to non-deterministic aspects of the GPU computing framework. All forecasts begin at 00UTC 20 June 2021.}
  \label{fig:graphcast_loss_10day}
\end{figure}

The evolution of the PNW loss as a function of time reveals how errors grow in the optimized forecasts relative to the control (Fig.~\ref{fig:graphcast_loss_10day}). 
The non-deterministic nature of JAX run on an NVIDIA A100 GPU yields an  ensemble for each experiment starting from identical initial conditions by simply repeating the forecast. Error in the control increases approximately exponentially with a doubling time of $\sim$1.4 days (estimated from linear regression). Error for the globally optimized solution grows similarly to the control over the first 5 days, but then declines such that by day 8 the error is comparable to day 3. Error then begins to increase again, but by 10 days the loss is $\sim$85\% less than the control, with an implied exponential average doubling time of $\sim$2.5 days (estimate from the 10-day loss). The PNW-optimized solution shows little growth in error until day 8, and at day 10 the loss is over $\sim$90\% less than the control, with an implied exponential average doubling time of $\sim$3.5 days (estimate from the 10-day loss).

While it is extraordinary that the initial conditions can be optimized to produce a nearly perfect 10-day forecast of the heatwave, two key questions remain: (1) to what extent do the initial conditions correct for model error rather than analysis error?, and (2) are the initial condition changes practically realizable with an operational analysis and forecasting system? We address the first question by producing forecasts with a different model, Pangu-Weather, initialized with the GraphCast-optimized initial conditions. We address the second question by analyzing the magnitude of the initial-condition differences from the ERA5, and by filtering the optimal initial-condition differences from the control forecast to eliminate small spatial scales. Together, these experiments reveal whether the improvements are due to infinitesimal changes at small scales (``butterflies'') or finite-amplitude changes at synoptic and larger scales, resolvable by the current observing system.

\subsection{Pangu-Weather Forecast with GraphCast-Optimized Inputs}

For the imperfect model experiments, we use the Pangu-Weather model \citep{bi2023}, which has been shown to have similar forecast skill to GraphCast \citep{rasp2023weatherbench}. The Pangu-Weather model architecture is distinctly different from GraphCast, as is the method of inference; we highlight the differences relevant to our experiments. For inputs, the models share isobaric levels and variables with the exception of vertical velocity and precipitation, which are not included in Pangu-Weather. The horizontal spatial resolution of Pangu-Weather is 0.25$^{\circ}$ as compared to 1$^{\circ}$ for the small version of GraphCast used here. In order to map the GraphCast-optimal initial condition to the Pangu-Weather grid, we interpolate by projecting each variable in the initial condition onto spherical harmonics, pad the result at small scales with zeros to the Pangu-Weather resolution, and backtransform to the  0.25$^{\circ}$ latitude--longitude grid. For inference, GraphCast employs an autoregressive procedure using two time levels separated by 6 hours to predict the next 6-hour step, whereas Pangu-Weather employs an autoregressive procedure using a single time step to predict the next 24-hour step. We use only the GraphCast-optimized input at 00 UTC 20 June 2021 for Pangu-Weather, while GraphCast uses this and the additional optimized input at 18 UTC 19 June 2021. In summary, the modeling frameworks differ significantly in resolution (both spatially and in time), input variables, network architecture, and inference procedure. Consequently, we believe these experiments provide a useful test for the contribution of GraphCast-specific model error that is encoded in the optimized initial condition.

\begin{figure}[ht!]
\begin{center}
  \noindent\includegraphics[width=.9\textwidth]{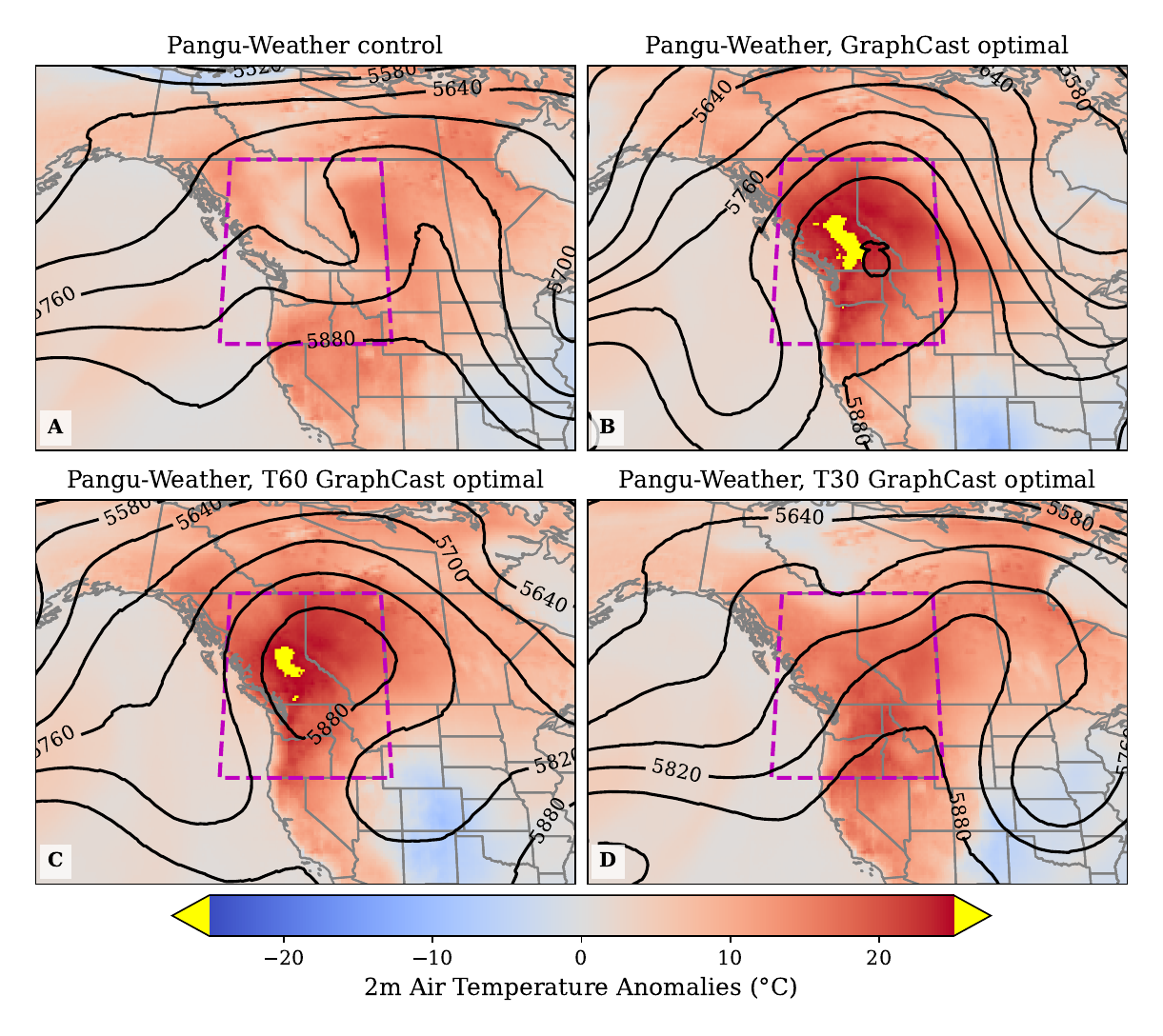}
\end{center}
  \caption{500 hPa geopotential height (contours) and 2m air temperature anomalies (colors) at 00UTC 30 June 2021 for A) Pangu-Weather 10-day control forecast, B) Pangu-Weather 10-day forecast initialized with the GraphCast PNW optimal initial condition C) as in B) but filtering the GraphCast optimal IC perturbations to T60 spherical harmonic truncation, and D) as in C) except for T30 GraphCast initial condition perturbations. All forecasts are initialized 00UTC 20 June 2021. Anomalies are relative to ERA5 June climatology 1979--2020, with anomalies greater than 25$^{\circ}$C shaded yellow.}
  \label{fig:pangu_10day}
\end{figure}

The Pangu-Weather control 10-day forecast shows a solution similar to the GraphCast control, with a 500 hPa trough over the PNW as compared to a ridge in the verification field (Fig.~\ref{fig:pangu_10day}A). The Pangu-Weather 10-day forecast initialized with the GraphCast-optimized initial conditions reveals a remarkable improvement over the control for both 500 hPa geopotential height and 2m air temperature anomalies (Fig.~\ref{fig:pangu_10day}B). Notable differences from the verification fields include a slightly weaker ridge in the Pangu-Weather forecast, and 2m air temperature anomalies that are too small, especially over eastern Washington and Oregon, by $\sim$3--5$^{\circ}$C; however, for a 10-day forecast of an extreme-outlier event, this is remarkably accurate. This result suggests that model error is not a significant contribution to the 10-day optimized initial condition, and that it largely captures analysis error in ERA5. 

Turning to the second key question--whether the optimal initial condition is realizable by an operational forecast system--we first note that these changes are of modest magnitude, similar to observation and analysis errors (Table 1). Specifically, the maximum absolute values in the global-optimum perturbations are $\sim$1~K in temperature, $\sim$1~ms$^{-1}$ in wind speed, $\sim$1~gkg$^{-1}$ in water vapor specific humidity, and $\sim$1~hPa in mean-sea-level pressure. Moreover, while the median values over the entire domain are small, they are not infinitesimal (i.e., not near floating point precision). Regional-optimum perturbation maxima and medians are within 20\% of globally-optimized values and tend to be smaller (not shown). In terms of the time--space evolution of forecast error changes in the 500 hPa geopotential height field, the optimal initial condition increases forecast error relative to ERA5 for lead times less than a day, followed by increasingly large improvements with lead time nearly everywhere (Fig. S2).

\begin{table}[h]
    \centering
    \begin{tabular}{l l l}
        \hline
        \textbf{Variable} & \textbf{Median Difference} & \textbf{Max Difference} \\
        \hline
        Geopotential Height & 1.3 m & 40.5 m \\
        Specific Humidity & 1.2e-03 g/kg & 2.6e-1 g/kg \\
        Temperature & 0.05 K & 0.81 K \\
        U Component of Wind & 0.04 m/s & 0.77 m/s \\
        V Component of Wind & 0.03 m/s & 0.53 m/s \\
        Vertical Velocity & 0.0003 m/s & 0.0111 m/s \\
        Mean Sea Level Pressure & 0.06 hPa & 0.71 hPa \\
        10m U Component of Wind & 0.02 m/s & 0.29 m/s \\
        10m V Component of Wind & 0.02 m/s & 0.26 m/s \\
        2m Temperature & 0.08 K & 0.98 K \\
        Total Precipitation (6 Hrly) & 6.9e-04 cm & 9.2e-03 cm\\
        \hline
    \end{tabular}
    \vskip 0.5truecm
    \caption{Median and maximum absolute differences between ERA5 and the 10-day globally optimized input for GraphCast variables shown in the left column.}
    \label{tab:median_max_diff}
\end{table}

In order to determine the contribution from small scales, we spectrally truncate the GraphCast PNW-optimized initial condition to T60 ($\sim3^{\circ}$ resolution), which eliminates over 99\% of the degrees of freedom in spectral space. The Pangu-Weather forecast from this truncated initial condition is very similar to that for the full optimal initial condition (Fig.~\ref{fig:pangu_10day}C). Truncating further to T30 ($\sim6^{\circ}$ resolution) results in a weaker ridge and temperature anomalies, but still a significant improvement over the control forecast (Fig.~\ref{fig:pangu_10day}D). These truncated-initial-condition experiments reveal that the important changes in the optimal initial condition are found on synoptic and planetary scales, and that the forecast differences are a smooth function of these truncated changes to the inputs. We conclude that these modest-amplitude, synoptic-scale, analysis errors may be resolvable by the current observational network. Improvements to ERA5-type initial conditions may be realized by better data assimilation systems, forecast models and, of course, improvements in the observing network.

\subsection{Globally-Optimized Inputs at Multiple Target Lead Times}

Given the dramatic increase in 10-day forecast skill from the optimal initial conditions for this case, we now briefly address forecasts for longer lead times. Given space limitations, we consider only one initial time, and will report a more general investigation of these questions in a separate paper. We consider 10 different optimization times from 2.5--25 days, starting from a different initialization time than considered previously, which demonstrates that the results for 10 day forecasts initialized 20 June are not peculiar to that time period.

\begin{figure}[ht!]
\begin{center}
  \noindent\includegraphics[width=.85\textwidth]{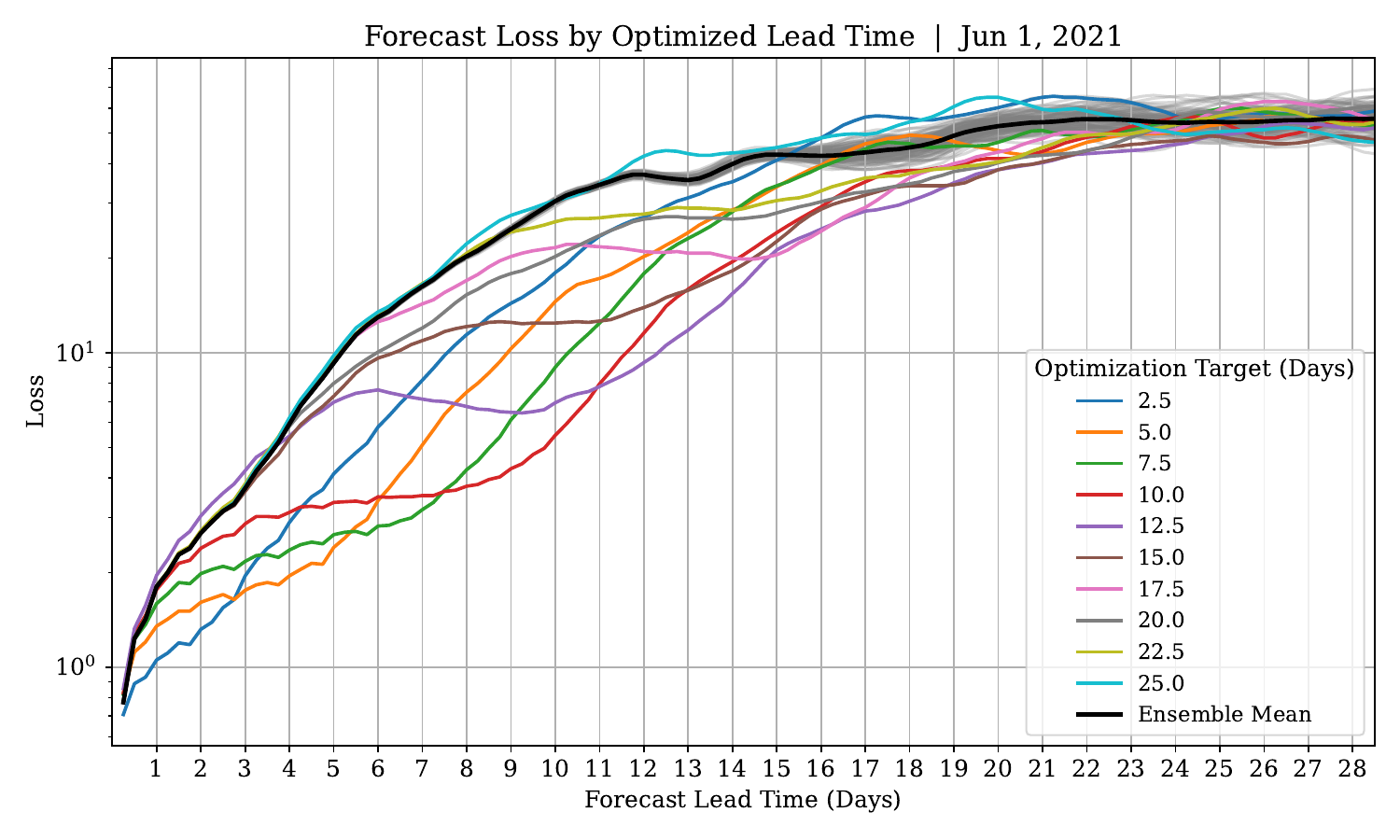}
\end{center}
  \caption{GraphCast instantaneous global loss (not time averaged) as a function of time starting 12 UTC 1 June 2021. A control ensemble forecast is shown in gray lines with the ensemble mean in black. All ensemble members have identical initial conditions given by the ERA5 analysis, and diverge due to non-deterministic aspects of the GPU computing framework. Color lines show forecasts optimized for lead times ranging from 2.5 to 25 days for a randomly chosen member of the control ensemble.}
  \label{fig:graphcast_29day}
\end{figure}

We consider here forecasts for 00 UTC 30 June 2021 starting on 12 UTC 1 June; that is, 28.5-day forecasts, again using control forecasts derived from GraphCast initialized with the ERA5 reanalysis. The control shows approximately exponential growth in forecast error from 1--7 days, followed by slower growth to saturation around 20 days (Fig.~\ref{fig:graphcast_29day}). The gray lines in Fig.~\ref{fig:graphcast_29day}) show the forecast trajectories for 100 different forecasts from identical ERA5 initial conditions with the black curve representing the ensemble mean. This ensemble provides an estimate of  whether solutions from the optimal initial conditions are meaningfully different from forecasts that diverge due solely to computational noise. 

Thin colored lines on the Fig.~\ref{fig:graphcast_29day} show results for forecasts from initial conditions optimized for different lead times for one of the randomly chosen control ensemble members. For example, the 10-day optimized forecast (red line) yields the most skillful forecast at 10 days (12UTC 11 June 2021), and it shows very little error growth until that time. Although forecast error grows rapidly after the optimization time, the forecast from this initial condition improves upon the control ensemble out to $\sim$22 days. After a period of initial growth, forecast errors for 12.5-day and 17.5-day optimization show error {\it decay} with lead time leading up to the optimal time. This suggests that the optimization procedure has identified nearby initial conditions that nonlinearly evolve on the attractor near the true trajectory, but with small departures that allow the solution to remain close to the true state at the optimization time. The least effective optimal in this regard is for the shortest optimization lead time of 2.5 days, which shows a delay in exponential growth, but is otherwise similar to the control. 

The long-time limit of the optimization procedure appears to be around 22.5 days, which improves upon the control even when allowing for an ensemble solution for the optimal initial condition (identical initial conditions; Fig. S3, yellow lines). After 22.5 days, the optimal perturbations yield solutions no different than the control ensemble, but this may be limited by the bfloat16 representation of floating-point numbers in JAX, which has only a 7 bit representation for the mantissa. For completeness, we note that the long-time limit for PNW optimals is also around 22.5 days (not shown).

\section{Discussion \&\ Conclusion\label{sec:conlusions}}

Sensitivity to initial conditions is a hallmark of atmospheric predictability, ultimately limiting the accuracy of long-term forecasts. Traditionally this sensitivity is measured using adjoint models to estimate the features in the initial conditions that grow optimally at a chosen lead time as measured in a chosen norm. Since adjoint models involve linearizing about a control forecast, applications are typically limited to a few days. Moreover, the technical challenges of coding the adjoint model are time consuming and involve approximations for derivatives for on/off processes. Deep-learning models provide the opportunity for a new approach to atmospheric predictability since they are networks of linear models connected by known, differentiable, nonlinear activation functions. Importantly, the frameworks used to train these models offer tools to compute gradients with respect to not just model parameters, but state variables as well. They also include algorithms by which to perform gradient-descent optimization. Here we define an algorithm that extends adjoint sensitivity to the fully nonlinear case to optimize initial conditions for forecast improvement at long lead times. This method's flexibility means that, in addition to the domain volumes considered here, input optimization may target specific variables, levels, timeframe, or location. We apply the method to the extreme heatwave over the Pacific Northwest in June 2021 using the GraphCast model and the JAX framework for automatic differentiation and optimization.

We find a $\sim$90\% reduction in 10-day forecast error for the PNW heatwave over a control forecast initialized from the ERA5 reanalysis. In theory the optimal initial conditions also encode corrections for model error, but similar forecast improvement from Pangu-Weather model forecasts starting from the GraphCast-optimized initial conditions suggests that the important perturbations are not specific to the forecast model. Truncating small scales from the optimal initial conditions reveals that the forecast improvements are due to finite-amplitude changes on synoptic and larger scales. The fact that the long-lead solutions vary smoothly in the truncation of the initial conditions suggests that, unlike physics models, the deep-learning models are not subject to noisy on/off processes (e.g., convective and microphysical parameterizations) that effectively add stochastic error to the forecast. Assessing the generality of this conjecture is an important direction for future research, especially in the context of interpreting differences in error growth between physics models and deep-learning models \citep{selz2023can}.

In addition to assessing the generality of the results presented here, future studies may also consider the joint optimization problem where both the model weights and the initial conditions are optimized. This could allow for flexible models that adapt optimally to a particular state and loss function, such as specific types of storms in a limited geographical area. Applications to data assimilation are also particularly compelling, as minimizing the loss from the misfit to observations \citep{hatfield2021building} should allow for long-window 4DVAR relative to the short windows currently used ($\sim$12 hours).

\section{Acknowledgements\label{sec:ack}}
We acknowledge high-performance computing support from the Casper cluster (doi.org/10.5065/qx9a-pg09) provided by NCAR's Computational and Information Systems Laboratory, sponsored by the National Science Foundation. The Copernicus Climate Data Store provided access to ERA5. We thank Dr. Chris Snyder (NCAR) for conversations related to nonlinear singular vectors and initial-condition error growth, and Mu Mu (Fudan University) for sharing his published papers on nonlinear adjoint methods. Trent Vonich was funded by the Dr. Heather Wilson STEM Fellowship. GJH acknowledges support from NSF award 2202526 and Heising-Simons Foundation award 2023-4715. 


%
%
%

%
%

%

\section{Open Research}
All ERA5 data used in this study is openly available from the Copernicus Data Store. All code required to operate GraphCast and Pangu-Weather can be found at the \url{https://github.com/google-deepmind/graphcast} and \url{https://github.com/198808xc/Pangu-Weather}, respectively. Optimal initial conditions will be posted to a public DOI when the paper is accepted for publication.







\bibliographystyle{unsrtnat}
\bibliography{vonich_hakim} 




\end{document}